\documentclass[onecolumn,eqsecnum,aps,showkeys,superscriptaddress, 11pt,pre]{revtex4}
\usepackage{epsfig}
\usepackage{color}
\usepackage{dcolumn}
\usepackage{xspace}
\usepackage{soul}
\usepackage{subfig}

\begin{document}
\title{Effect of magnetic nanoparticles on the nematic-smectic-A phase 
transition}
\author{Prabir K. Mukherjee}
\affiliation{Department of Physics, Government College of Engineering
and Textile Technology, 12 William Carey Road, Serampore, Hooghly-712201,
India}
\email{pkmuk1966@gmail.com}
\author{Amit K Chattopadhyay}
\affiliation{
Aston University, System Analytics Research Institute, Birmingham, B4 7ET, UK}
\email{a.k.chattopadhyay@aston.ac.uk}
\date{\today}
\begin{abstract}
\centerline{Abstract}
\noindent
Recent experiments on {\it mixed} liquid crystals have highlighted the hugely significant role of ferromagnetic nanoparticle impurities in defining the nematic-smectic-A phase transition point. Structured around a Flory-Huggins free energy of isotropic mixing and Landau-de Gennes free energy, this article presents a phenomenological mean-field model that quantifies the role of such impurities in analyzing thermodynamic phases, in a mixture of thermotropic smectic liquid crystal and ferromagnetic nanoparticles.
First we discuss the impact of ferromagnetic nanoparticles on the isotropic-ferronematic and ferronematic-ferrosmectic phase 
transitions and their transition temperatures.  
This is followed by plotting and
discussing various topologies in the phase diagrams.
Our model results indicate that there exists a critical concentration of nanoparticle impurities for which the second order N-SmA transition becomes first order at a tricritical point. Calculations based on this model show remarkable agreement with
experiment.
\end{abstract}
\maketitle
\newpage

\section{Introduction}

In recent years, many experiments have found that liquid crystals doped with 
dispersive materials e.g.
carbon nanotubes, silica microbeads,
nanoparticles and various collides 
exhibit remarkable
new physical phenomena. Experiments have shown that
nanoparticles and ferroelectric nanoparticles can greatly
enhance the physical properties of nematic and smectic
liquid crystals \cite{reznikov,li1,milkulko,blach,cook,reshetnyak,kurochkim,ouskova,gharde,basu,rzoska}.

The mixture of nematic and smectic liquid crystals and magnetic
nanoparticles is known as a ferronematic and ferrosmectic.
A number of experimental studies
\cite{miranda,potocova,cordo3,fabre,ramos,ponsinet1,ponsinet2,spoliansky1,spoliansky2} are devoted to
study of ferrosmectic phase in both in thermotropic and lyotropic liquid crystals. The most essential feature of the ferrosmectics is that their structure aligns parallel to an external magnetic field.
Potocova et al. \cite{potocova} studied the structural instabilities 
of the ferronematic and ferrosmectic phases prepared from 8CB 
(CH$_3$(CH$_2$)$_7$(C$_6$H$_4$)$_2$CN). Martínez-Miranda et al. \cite{miranda}
studied how the surface 
coating interacts with the liquid crystal in conjunction with the ferromagnetic nanoparticles (FNP). They have
found out that depending on the surface coating the interaction of the 
ferromagnetic nanoparticles with the liquid
crystal varies. 
Cordoyiannis et al. \cite{cordo3} experimentally studied the impact of magnetic
nanoparticles on the isotropic-nematic (I-N) and nematic-Semectic-A (N-SmA) 
phase transitions of 8CB.
This work shows that the I-N
transition remains weakly first order even in the presence of FNP. 
For the N-SmA phase transition the results of this work show
a crossover from anisotropic criticality towards tricriticality.

A large number of theoretical works \cite{lopatina1,lopatina2,soule,popa1,kralj,muk1,muk2} have been 
carried out to describe the effect of anisotropic nanoparticles, carbon nanotubes and ferroelectric nanoparticles in nematic  and smectic liquid crystals.
However, very few or practically no theoretical work has been attempted to
explain the phase behavior of the N-SmA phase transition in thermotropic
liquid crystals (TLC) in the mixture of FNP. 
Only one theoretical work \cite{muk3} on ferrosmectic phase in LLC is 
available in the literature but that is not quantitatively explicit either. 
Thus it is interesting to see how the ferromagnetic nanoparticles 
influence the character of the N-SmA phase transition in TLC. 
Based on this core question, here we to develop a phenomenological model
structured around the Flory-Huggins theory \cite{flory} to discuss the I-N and N-SmA phase transitions in
the in the mixture of FNP. 

\section{Model}

In this section, we use the combination of Flory-Huggins theory and Landau-de 
Gennes theory for the binary mixture of calamitic SmA liquid crystal and
ferromagnetic nanoparticles. First we describe the order parameters necessary
in the model free energy.
The smectic-A phase has both the orientational 
and translational ordering. The nematic order parameter,
originally proposed by de Gennes \cite{degennes}, is a symmetric, traceless
tensor described by $Q_{ij}=\frac S2(3n_in_j-\delta _{ij})$, where $n_i$ are unit vectors specifying the preferred orientation of the primary molecular axes, also called directors. The quantity 
$S$ defines the strength of the nematic ordering . The layering in the
SmA phase is characterized \cite{degennes} by the order parameter 
$\psi (\bf{r})=\psi _0\exp (-i\Psi )$, which is a complex scalar quantity whose
modulus $\psi _0$ is defined as
the amplitude of a one dimensional density wave characterized by the phase
$\Psi$.
The magnetic order is described by the magnetization 
${\bf{M}}=M \hat{\bf{m}}$ such that $|M|=0$ in paramagnetic state and 
$|M| \neq 0$ in the ferromagnetic state.
Thus we use $Q_{ij}$, $\psi$, and $\bf{M}$ 
as order parameters necessary for the description of the I-N and N-SmA
phase transitions in the mixture of FNP.

The total free energy per unit volume of the mixture can be written as
\begin{equation}
f=f_{\text{mix}}+f_{NP}+F_{LC}+f_{\text{int}},
\label{free1}
\end{equation}
where $f_{mix}$ is the free energy mixing of isotropic liquids; $f_{NP}$ 
describes the contribution of the FNP dispersed in liquid crystal, $f_{LC}$ 
represents the free energy of SmA ordering of liquid crystals and $f_{int}$ describes
the coupling between FNP and SmA ordering, respectively.

The isotropic mixing free energy density may be approximated in terms of 
the Flory-Huggins theory \cite{flory} 
\begin{equation}
\frac{f_{\text{mix}}}{k_BT}=\phi \ln\phi+(1-\phi)\ln(1-\phi)+\chi \phi(1-\phi) +\frac{\lambda}{2} {\bigg( {\bf \nabla} \phi \bigg)}^2,
\label{free2}
\end{equation}
where $k_B$ is the Boltzmann constant and $T$ is absolute temperature. 
$\phi$ and $(1-\phi)$ describe the volume fractions of FNP and SmA liquid 
crystals. $\frac{\lambda}{2} {\bigg( {\bf \nabla} \phi \bigg)}^2$ is the Ginzburg-Landau term. $\chi$ is a temperature-dependent function which can be defined as $\chi=A+\frac {B}{T}$ is known as the Flory-Huggins interaction 
parameter with $A$ and $B$ are constants.

The contribution of the FNP free energy density can be expressed as
\begin{equation}
f_{NP}=\phi\left[\frac 12p{\bf M}^2
+\frac 14q{\bf M}^4\right]
\label{free3}
\end{equation}
The material parameter $p$ can be assumed as $p=p_0(T-T_3^*)$. $p_0$ is a 
positive constant and $T_3^*$ is the virtual
transition temperature. 
We assume $q>0$ for the stability of the free energy.

The SmA free energy density can be expressed as
\begin{eqnarray}
f_{LC}&=&(1-\phi) \bigg[\frac 13aQ_{ij}Q_{ij}-\frac 49bQ_{ij}Q_{jk}Q_{ki}
+\frac 19c\left(Q_{ij}Q_{ij}\right) ^2
\nonumber \\
&&+\frac 12\alpha \left| \psi \right| ^2
+\frac 14\beta \left| \psi \right| ^4-\frac 12 \delta_{ij} Q_{ij} \left| \psi \right| ^2 \bigg]
\label{free4}
\end{eqnarray}
$a$ and $\alpha$ can be assumed as
$a=a_0(T-T_1^{*})$ and $\alpha=\alpha_0(T-T_2^*)$,
$a_0>0$
and $\alpha_0>0$.
$T_{1}^{*}$ and $T_2^{*}$ are the virtual
transition temperatures.
We choose $c>0$, $b>0$ and $\beta>0$ for the stability of the free energy
density (\ref{free4}). Equations (2.2)-(2.5) are all tacitly structured around the standard symmetricity argument.

The contribution to free energy density due the interactions is written as
\begin{equation}
f_{\text{int}}=-\phi(1-\phi)\left[\frac 12\gamma M_iM_jQ_{ij} 
+\frac 13\eta_1 {\bf M}^2Q_{ij}Q_{ij}
+\frac 12\eta_2M_iM_kQ_{ij}Q_{kj}+\frac 12 \omega {\bf M}^2\left| \psi
\right| ^2\right]
\label{free5}
\end{equation}
The parameters $\delta$, $\gamma$, $\eta_{1,2}$, and
 $\omega$
are coupling constants.  $\delta $ is chosen positive to favor the
smectic-A   
phase over the nematic phases The positive values
of $\gamma$, $\eta_{1,2}$, and $\omega$ ensures the ferromagnetic 
order induced by the nematic order and translational order. 

Following Pleiner et al. \cite{pleiner} we consider  
the ordering directions between $\hat{\bf {n}}$ and 
$\hat{\bf{m}}$ make an angle $\theta_f$ i.e. $\hat{\bf {n}}\cdot \hat{\bf {m}}=
\cos\theta_f$. Then the total free energy density (\ref{free1}) leads to
\begin{eqnarray}
f &=&k_BT\bigg[\phi \ln\phi+(1-\phi) \ln(1-\phi)+\chi \phi(1-\phi) + \frac{\lambda}{2}{\bigg( {\bf  \nabla} \phi \bigg) }^2\bigg]\nonumber \\
&&+(1-\phi)\bigg[\frac 12aS^2-\frac 13bS^3+\frac 14cS^4+\frac 12\alpha \psi_0^2
+\frac 14\beta \psi _0^4
-\frac 12 \delta \psi_0^2 S\bigg] 
+\phi \left[\frac 12p M^2
+\frac 14q M^4\right] \nonumber \\
&&-\phi(1-\phi)\bigg[\frac 14\gamma M^2S(3\cos^2\theta_f-1) 
+\frac 12\eta_1 M^2S^2 
+\frac 18\eta_2M^2S^2(3\cos^2\theta_f+1) \nonumber \\
&&+\frac 12 \omega M^2\psi_0^2\bigg]
\label{free6}
\end{eqnarray}
\noindent
Minimization of Eq. (\ref{free6}) with respect to $S$, $\psi_0$, $M$ and 
$\theta_f$ yields the following five stable solutions excluding the ferromagnetic 
state:

\begin{enumerate}
\item[(I)] Isotropic phase (I):
$S=0$, $\psi_0=0$, $M=0$, $\theta_f=0$.
This phase exists for $a>0$, $\alpha >0$, and $p>0$.

\item[(II)] Nematic phase (N): $S\neq 0$, $M=0$, $\psi_0=0$, $\theta_f=0$. 
This phase exists for $a<0$, $\alpha-\delta S >0$, $p-(1-\phi)\gamma^{\prime}S-
(1-\phi)\eta^{\prime}S^2>0$.

\item[(III)] Smectic-A phase (SmA): $S\neq 0$, 
$\psi _0\neq 0$, $M=0$, $\theta_f=0$.

This phase exists for $a<0$, $\alpha-\delta S<0$,
and 
$p-(1-\phi)\gamma^{\prime}S-(1-\phi)\eta^{\prime}S^2 + \omega \psi_0^2>0$.

\item[(IV)] Ferronematic phase (FN): $S\neq 0$, $\psi_0=0$, $M \neq 0$, 
$\theta_f=0$ or $\theta_f=\frac {\pi}{2}$. 
This phase exists for $a<0$, $p-(1-\phi)\gamma^{\prime} S-(1-\phi)\eta^{\prime}S^2<0$,
$\alpha-\delta S>0$. 

\item[(V)] Ferrosmectic phase (FSmA): $S\neq 0$, $\psi_0 \neq0$, $M \neq 0$, 
$\theta_f=0$ or $\theta_f=\frac {\pi}{2}$.
This
phase exists for $a<0$, $p-(1-\phi)\gamma^{\prime} S-(1-\phi)\eta^{\prime}S^2 + \omega \psi_0^2<0$,
$\alpha-\delta S<0$.
\end{enumerate}

In the description above, we have used $\eta' = \eta_1 + \frac{\eta_2}{4} (3 \cos^2 \theta_f +1)$ and $\gamma' = \frac 12(3\cos^2 \theta_f -1) \gamma$. For the specific cases in hand,
$\gamma^{\prime}=\gamma (or -\frac {\gamma}{2})$ for $\theta_f=0$ 
(or $\frac {\pi}{2}$) and $\eta^{\prime}=(\eta_1+\eta_2)$ 
(or $(\eta_1+\eta_2/2)$)
for $\theta_f=0$ 
(or $\frac {\pi}{2})$.

The necessary conditions for the four different phases to be
stable (Hessian determinant) are given below:
\begin{eqnarray}
\frac{\partial ^2f}{\partial S^2}>0, 
, \frac{\partial ^2f}{\partial M^2}
>0, \frac{\partial ^2f}{\partial \psi_0^2}>0, \nonumber \\
\frac{\partial ^2f}{\partial u^2}\cdot \frac{\partial ^2f}{\partial v^2}
-\left( \frac{\partial ^2f}{\partial u\partial v}\right) ^2 &>&0,  \nonumber
\label{stab}
\end{eqnarray}
where $u, v\in \{S,\psi_0,M\}$.

Now it is clear from the above solutions that
I-N, I-SmA, I-FN, I-FSmA,  
N-SmA, N-FN, FN-FSmA, FN-SmA, SmA-FSmA phase transitions are possible. 
I-N, I-SmA, I-FN, I-FSmA phase transitions must always be  
first order because of
the cubic invariant $b$ in the free energy expansion (\ref{free6}). Other phase transitions can 
be first or second order depending on the  
concentration of FNP.
In the following we will discuss only the I-FN and FN-FSmA phase transitions 
which is observed experimentally.

\subsection{I-FN phase transition}
In order to ensure the stability of the FN phase, we require
\begin{equation}
a^*-2b^*S+3c^*S^2>0,  \label{stab1a}
\end{equation}
\begin{equation}
p-(1-\phi)(\gamma^{\prime}S+\eta^{\prime}S^2)<0,   \label{stab2a}
\end{equation}
\begin{equation}
a_1-2b_1S+3c_1S^2>0,   \label{stab4}
\end{equation}
where

$a^*=a-\phi(1-\phi)\frac{\gamma^{\prime 2}}{2q}+\phi\frac {p\eta^{\prime}}
{q}$, $b^*=b+\phi(1-\phi)\frac {3\gamma^{\prime}\eta^{\prime}}{2q}$,
$c^*=c-\phi(1-\phi)\frac {\eta^{\prime 2}}{q}$,
$a_1=a^*-\phi(1-\phi)\frac{\gamma^{\prime 2}}{2q}$,
$b_1=b^*+\phi(1-\phi)\frac {\gamma^{\prime}\eta^{\prime}}{q}$,
$c_1=c^*-\phi(1-\phi)\frac {2\eta^{\prime 2}}{3q}$.

The renormalized coefficients show that the Landau coefficients $a^*$, $b^*$
and $c^*$ change with change of the concentration of FNP.

Now the free energy density near the I-FN phase transition can be expressed as
\begin{eqnarray}
f &=&k_BT\left[\phi \ln\phi+(1-\phi) \ln(1-\phi)+\chi \phi(1-\phi) +\frac{\lambda}{2} {\bigg( {\bf \nabla} \phi \bigg)}^2\right]\nonumber \\
&&+(1-\phi)\left[\frac 12aS^2-\frac 13bS^3+\frac 14cS^4\right]
+\phi\left[\frac 12p M^2
+\frac 14q M^4\right] \nonumber \\
&&-\phi(1-\phi)\left[\frac 12\gamma^{\prime} M^2S 
+\frac 12\eta^{\prime} M^2S^2 \right]
\label{free6a}
\end{eqnarray}
The value of the magnetization in the FN phase can be expressed as
\begin{equation}
M^2=-\frac {1}{q}(p-(1-\phi)(\gamma^{\prime}S+\eta^{\prime}S^2)),
\label{magn1}
\end{equation}
where the value of $S$ in the FN
phase can be calculated from the equations
\begin{equation}
\phi\frac {p\gamma^{\prime}}{2q}+a^*S-
b^*S^2+c^*S^3=0
\label{orien}
\end{equation}

The temperature variation of the order parameter
($S$) for pure sample and a fixed concentration of FNP in the FN phase is shown in Fig.1.
This is done for a set of
phenomenological parameters for which the direct I-N and I-FN phase transitions
are possible. Figure 1 shows that I-FN transition temperature and the jump 
of the order parameter $S$ decrease with the increase of the concentration of
FNP. For a fixed set of parameter values,  we find for pure sample 
$S_{I-N}=0.4$ and $T_{I-N}=313.36K$. For the volume fraction $\phi=0.02$, 
we find $T_{I-FN}=312.51$ are $S_{I-FN}=0.31$. 
The low value of
$S_{I-FN}$ indicates the weakly first order character of the
I-FN phase transition. Thus the I-FN transition is still a weakly first 
order transition even in the mixture of FNP. The present analysis completely agree with experimental 
results of Cordoyiannis et al. \cite{cordo3}.

The substitution of $M$ from Eq. (\ref{magn1}) into Eq. (\ref{free6a}), we
get 
\begin{eqnarray}
f_{NA} &=&k_BT\bigg[\phi \ln\phi+(1-\phi) \ln(1-\phi)+\chi \phi(1-\phi) + \frac{\lambda}{2} {\bigg( {\bf \nabla} \phi \bigg)}^2\bigg]\nonumber \\
&&+(1-\phi)\left[\phi\frac {p\gamma^{\prime}}{2q}S+\frac 12a^*S^2-\frac 13b^*S^3+\frac 14c^*S^4\right]
\label{free6b}
\end{eqnarray}
The free energy density (\ref{free6b}) describes the I-FN phase transition. 
At the dimensional level, the Ginzburg-Landau free energy term \big[$\frac{\lambda}{2} {\big( {\bf \nabla} \phi \big)}^2$\big] effectively renormalizes the value of the parameter $\chi$ by rescaling the quadratic $\phi^2$-term with a $\frac{\lambda}{L^2}$ type term, where $L$ is the typical scaling length of the system. Beyond the mean-field level, the $\frac{1}{L^2}$-dependence will further renormalize the spatial correlation function.
The cubic coefficient $b^*$ in the free energy density (\ref{free6b}) shows that
the I-FN phase transition must always be first order in mean field
approximation. Lower the value of $b^*$, weakly the first order character of
the I-FN phase transition. 

The conditions for the first order I-FN phase transition can
be obtained as
\begin{equation}
f_{FN}(S)=F_{0}(T), f_{FN}^{\prime}(S)=0, f_{FN}^{\prime \prime}(S)\ge 0
\label{cond1}
\end{equation}
The conditions for phase equilibrium require that the chemical potentials 
in the isotropic and FN phases are equivalent i.e $\mu_{iso}=\mu_{FN}$. 

The FN phase appears only for $a^{*}<0$ $i.e.$
\begin{equation}
T < T_{I-FN}^*+
\frac {\phi(1-\phi)\gamma^{\prime 2}}{2qa_0^*}
\label{ntemp1}
\end{equation}
where
$T_{I-FN}^*=\frac{a_0T_1^*+\frac{\phi\eta^{\prime}p_0T_f}{q}}{a_0^*}$,

$a_0^*=a_0+\frac{\phi\eta^{\prime}p_0}{q}$.

From Eq. (\ref{ntemp1}) we observe the decrease of the I-FN transition temperature 
with the increase of the concentration of FNP as
\begin{equation}
\Delta T_{I-FN}=\frac {\phi(1-\phi)\gamma^{\prime 2}}{2qa_0^*}
\label{ntemp2}
\end{equation}

\subsection{FN-FSmA phase transition}

We now discuss the FN-FSmA phase transition.
In order to ensure the stability of the FSmA phase we require
\begin{equation}
a^{**}-2b^{**}S+3c^{**}S^2>0,  \label{stab3a}
\end{equation}
\begin{equation}
\alpha^*-\delta^*S-\frac {\phi(1-\phi)\omega \eta^{\prime}}{q}S^2<0,
\label{stab3b}
\end{equation}
\begin{equation}
p^*-(1-\phi)\gamma^{*}S-(1-\phi)\eta^{*}S^2<0.
\label{stab3c}
\end{equation}
\begin{equation}
a_2-2b_2S+2c_2S^2>0,   \label{stab3d}
\end{equation}
\begin{equation}
\omega<\sqrt{q\beta/\phi(1-\phi)}
\label{stab3e}
\end{equation}
where

$a^{**}=a-\frac {\delta \delta^*}{2\beta^*}-\frac {\phi(1-\phi)\gamma^*\gamma^{\prime}}{2q}+\frac {\phi\eta^{\prime}p^*}
{q}$, $b^{**}=b+\frac{\phi(1-\phi)\delta\omega\eta^{\prime}}{2q\beta^*}+\frac {\phi(1-\phi)\gamma^{*}\eta^{\prime}}{q}+\frac{\phi(1-\phi)\gamma^{\prime}\eta^{*}}{q}$,
$c^{**}=c-\frac {\phi(1-\phi)\eta^{\prime}\eta^*}{q}$,
$a_2=a^{**}-\frac {\phi(1-\phi)\gamma^{\prime 2}}{2q}$,
$b_2=b^{**}+\frac{\phi(1-\phi)\gamma^{\prime}\eta^{\prime}}{q}$,
$c_2=c^{**}-\frac{\phi(1-\phi)2\eta^{\prime 2}}{3q}$, 
$\alpha^*=\alpha+\frac{\phi\omega p}{q}$,
$\delta^*=\delta+
\frac {\phi(1-\phi)\omega\gamma^{\prime}}{q}$, $\beta^*=\beta-\frac {\phi(1-\phi)\omega^2}{q}$,
$p^*=p+\frac {(1-\phi)\omega\alpha^*}{\beta^*}$, $\eta^{*}=\eta^{\prime}+\frac {\phi(1-\phi)\omega
\eta^{\prime}}{q\beta^*}$, $\gamma^{*}=\gamma^{\prime}+\frac {\omega\delta^*}{\beta^*}$.

The above renormalized coefficients show that the interaction parameters
$\delta$, $\eta^{\prime}$ and $\gamma^{\prime}$ change with the change of
the concentration of FNP.

The values of the smectic ordering
and the magnetization 
in the FSmA phase can be expressed as
\begin{eqnarray}
\psi _0^2 &=& -\frac{1}{\beta^*}\bigg(\alpha^*-\delta^*S-\frac {\phi(1-\phi)\omega \eta^{\prime}}{q}S^2\bigg),
\label{smec} 
\end{eqnarray}
\begin{equation}
M^2=-\frac {1}{q}\bigg(p^*-(1-\phi)(\gamma^{*}S+\eta^{*}S^2)\bigg). 
\label{magn}
\end{equation}
where the values of $S$ in the FSmA
phase can be calculated from the equation
\begin{equation}
\frac{\delta\alpha^*}{2\beta^*}+\frac {\phi\gamma^{\prime}p^*}{2q}+a^{**}S-
b^{**}S^2+c^{**}S^3=0.
\label{orien1}
\end{equation}
Equation (\ref{orien1}) is a cubic equation that admits of an exact solution (Cardan's method) involving the parameters involved.

It is clear from Eq. (\ref{smec}) and Eq. (\ref{magn}) that a nonzero real value of $\psi_0$ and $M$
exist only when 
$\alpha^*-\delta^*S-\frac{\phi(1-\phi) \omega \eta^{\prime}}{q} S^2<0$ and 
$p^*-(1-\phi)(\gamma^{*}S+\eta^{*}S^2)<0$.
Since there is a small temperature range where
$\alpha >0$, $\delta>0$, $\gamma^{\prime}>0$ and $\eta^{\prime}<0$ in this region.

The substitution of $\psi_0$ and $M$ from Eqs. (\ref{smec}) and (\ref{magn}) into Eq. (\ref{free6}) gives
\begin{eqnarray}
f_{FSmA} &=&k_BT\bigg[\phi \ln\phi+(1-\phi) \ln(1-\phi)+\chi \phi\bigg(1-\frac{\lambda}{2L^2} \phi\bigg)\bigg]\nonumber \\
&&+(1-\phi)\left[\frac{\delta\alpha^*}{2\beta^*}S+\frac {\phi\gamma^{\prime}p^*}{2q}S+\frac 12a^{**}S^2-\frac 13
b^{**}S^3+\frac 14c^{**}S^4\right]
\label{free6c}
\end{eqnarray}
The free energy density (\ref{free6c}) describes the FN-FSmA phase transition.
The cubic coefficient $b^{**}$ in the free energy density (\ref{free5}) shows that
the FN-FSmA phase transition must be first order in mean field
approximation in the mixture of FNP. 

\begin{figure}[!ht]
     \subfloat[$S$ vs $T$ for $\theta_f=0$, $\phi=0$\label{subfig1:fig4}]{%
       \includegraphics[width=0.3\textwidth,angle=270]{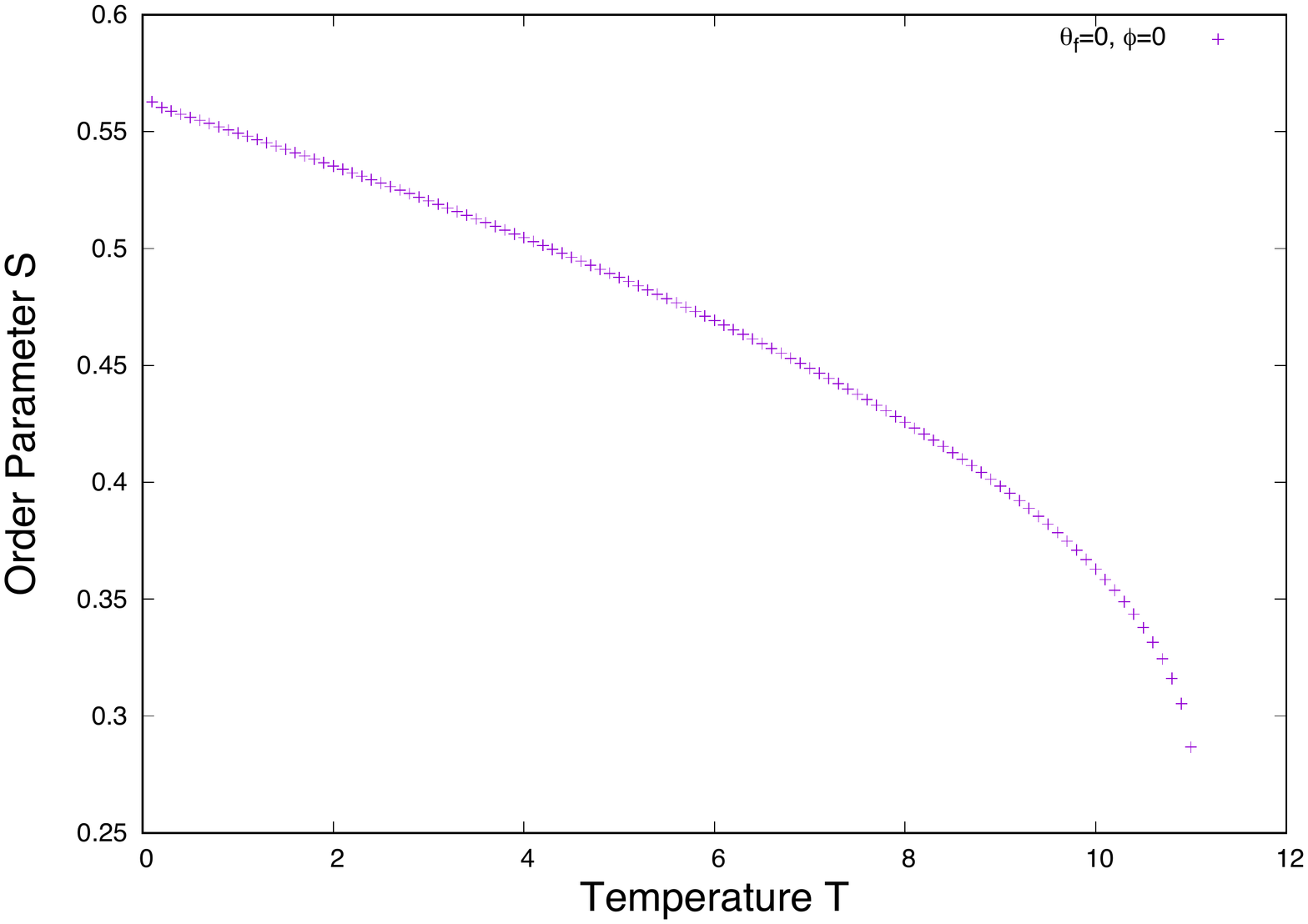}
     }
     \hfill
     \subfloat[$\psi_0^2$ vs $T$ for $\theta_f=0$, $\phi=0$\label{subfig2:fig4}]{%
       \includegraphics[width=0.3\textwidth,angle=270]{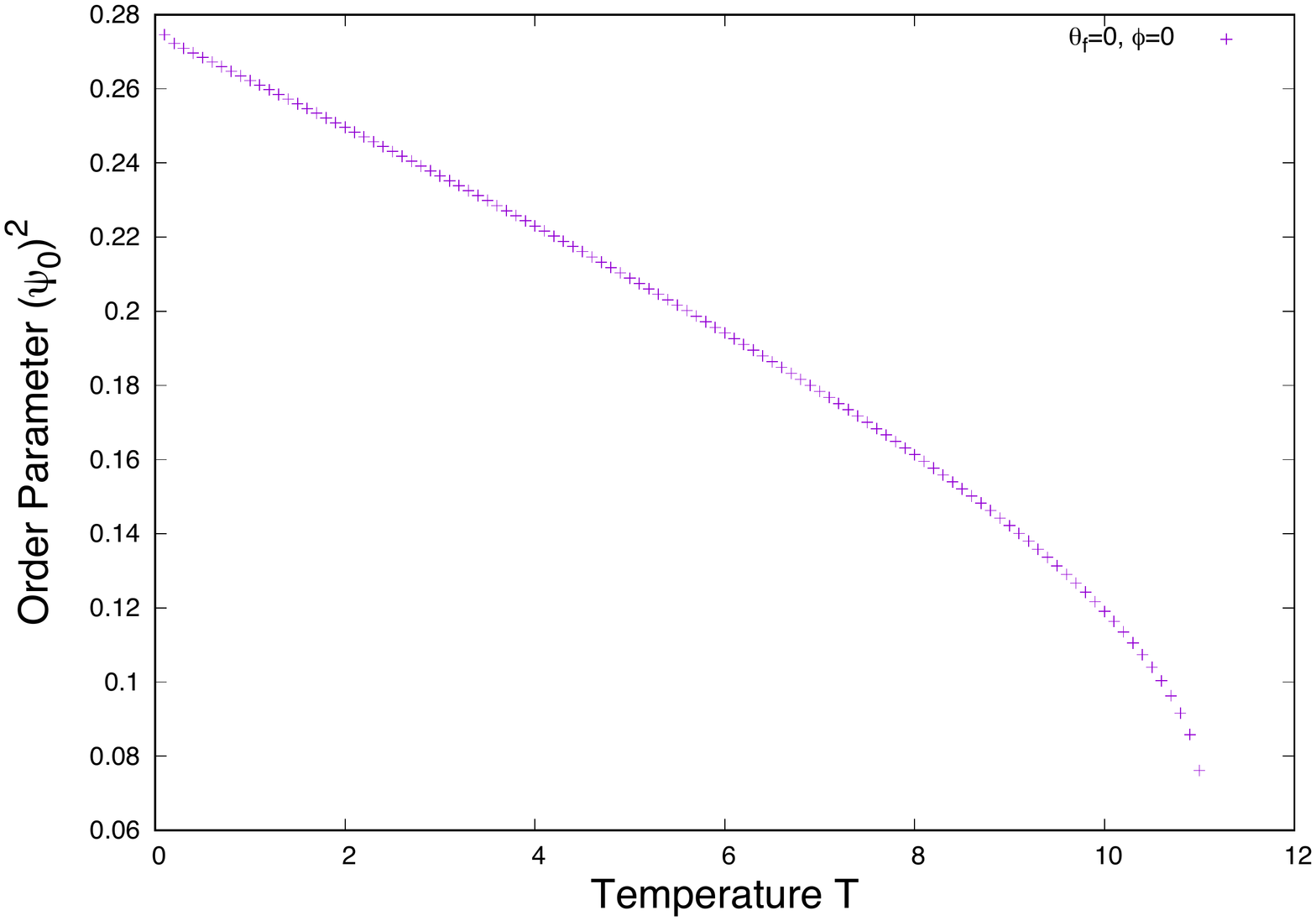}
     }
     \caption{$S$ and $\psi_0^2$ plotted against temperature $T$; show continuous decaying trend for $\phi=0$ with very little dependence on the $\theta_f$ values.}
     \label{fig1}
   \end{figure}

\begin{figure}[!ht]
     \subfloat[$S$ vs $T$ for $\theta_f=\pi/2$, $\phi=0$\label{subfig1:fig4}]{%
       \includegraphics[width=0.3\textwidth,angle=270]{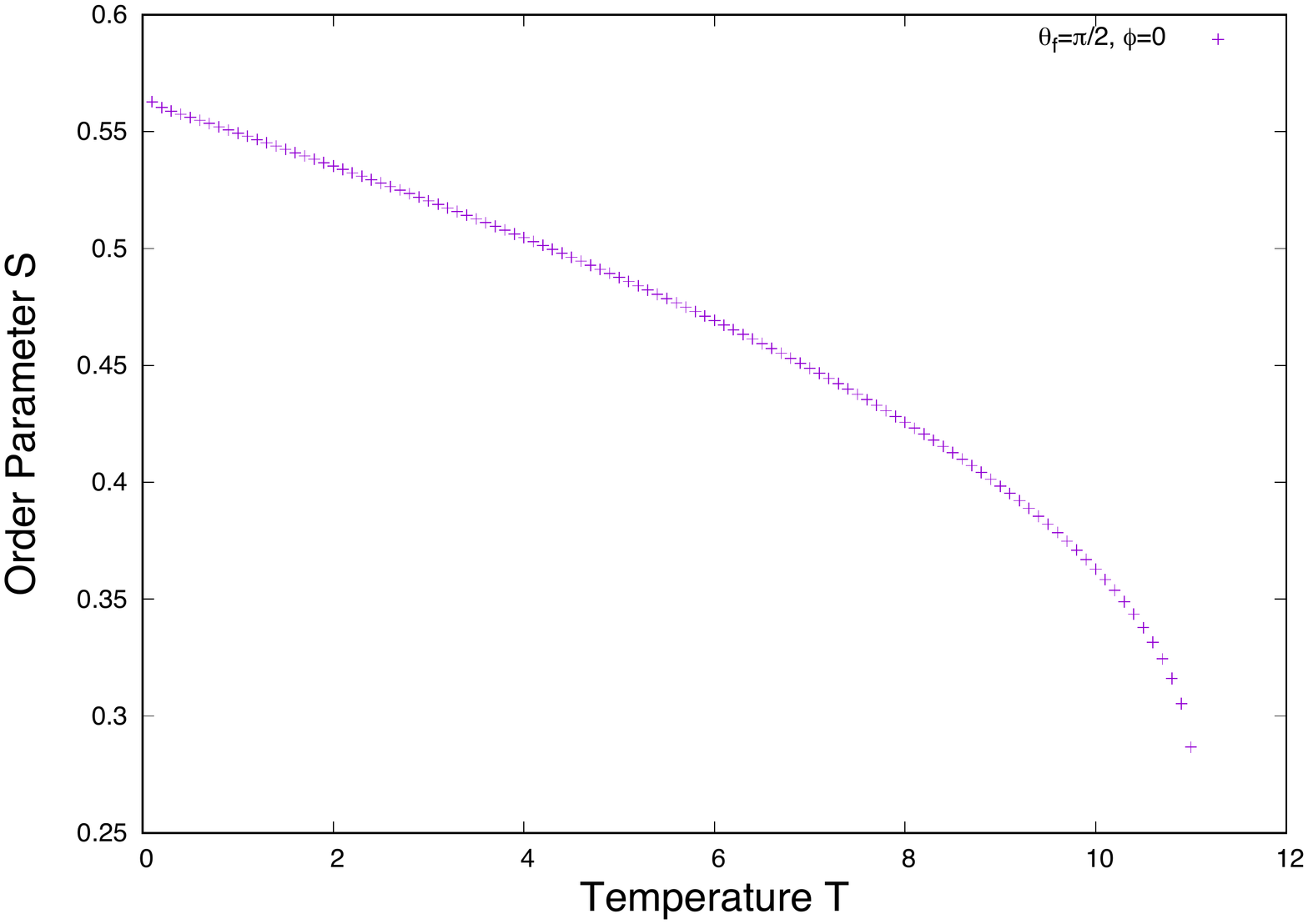}
     }
     \hfill
     \subfloat[$\psi_0^2$ vs $T$ for $\theta_f=\pi/2$, $\phi=0$\label{subfig2:fig4}]{%
       \includegraphics[width=0.3\textwidth,angle=270]{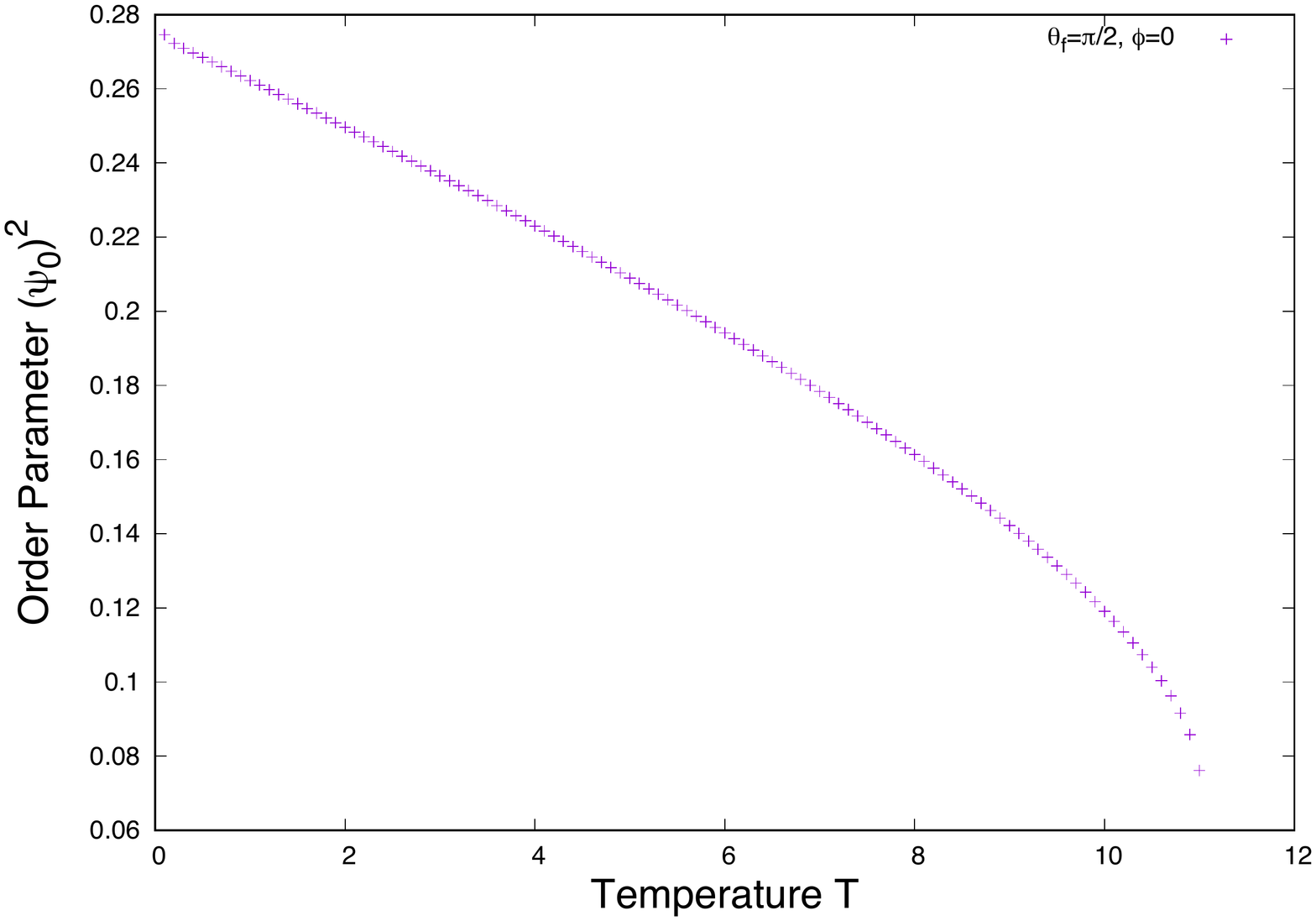}
     }
     \caption{$S$ and $\psi_0^2$ plotted against temperature $T$; show continuous decaying trend for $\phi=0$ with very little dependence on the $\theta_f$ values.}
     \label{fig2}
   \end{figure}

\begin{figure}[!ht]
     \subfloat[$S$ vs $T$ for $\theta_f=0$, $\phi=0.02$\label{subfig1:fig4}]{%
       \includegraphics[width=0.3\textwidth,angle=270]{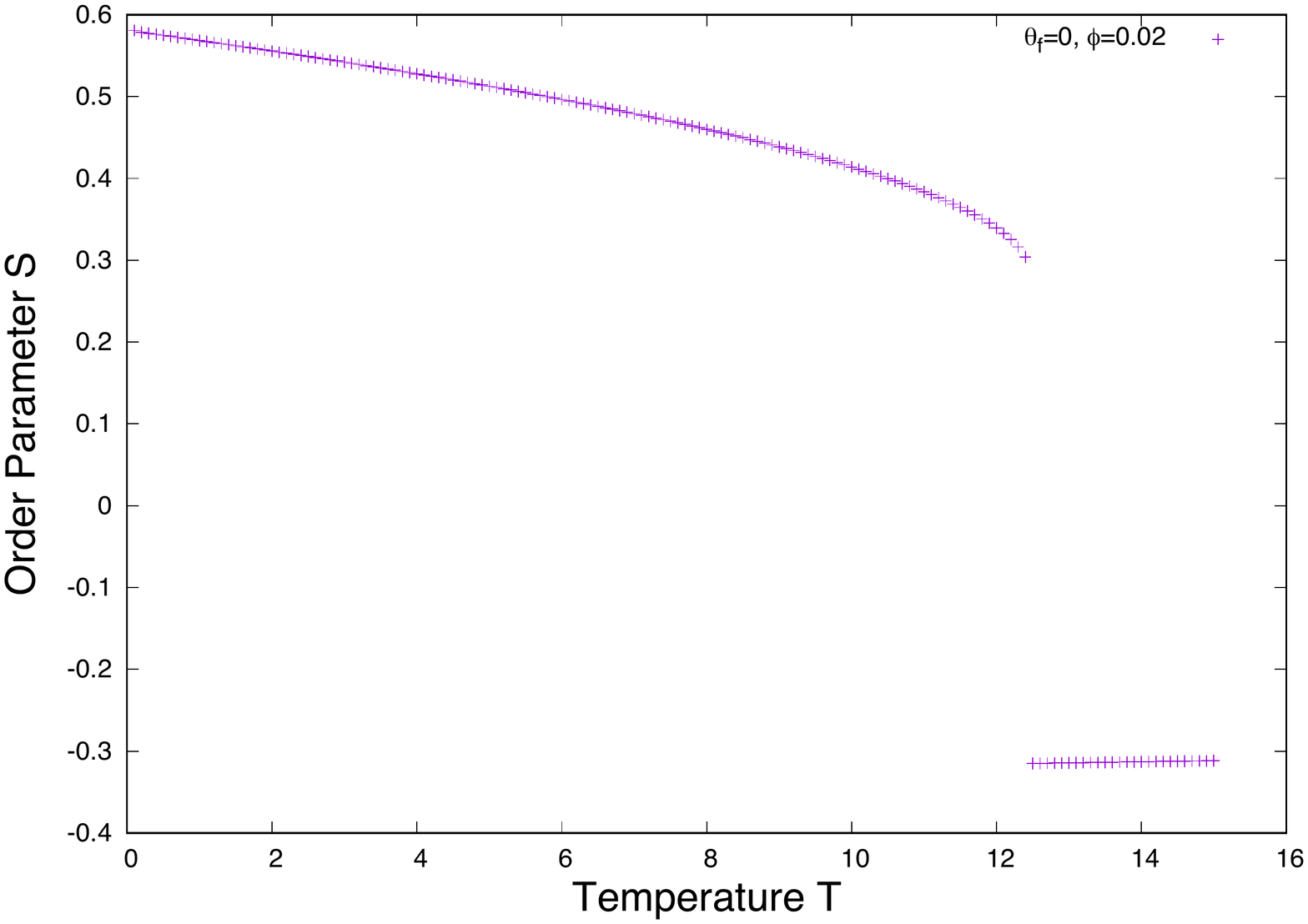}
     }
     \hfill
     \subfloat[$\psi_0^2$ vs $T$ for $\theta_f=0$, $\phi=0.02$\label{subfig2:fig4}]{%
       \includegraphics[width=0.3\textwidth,angle=270]{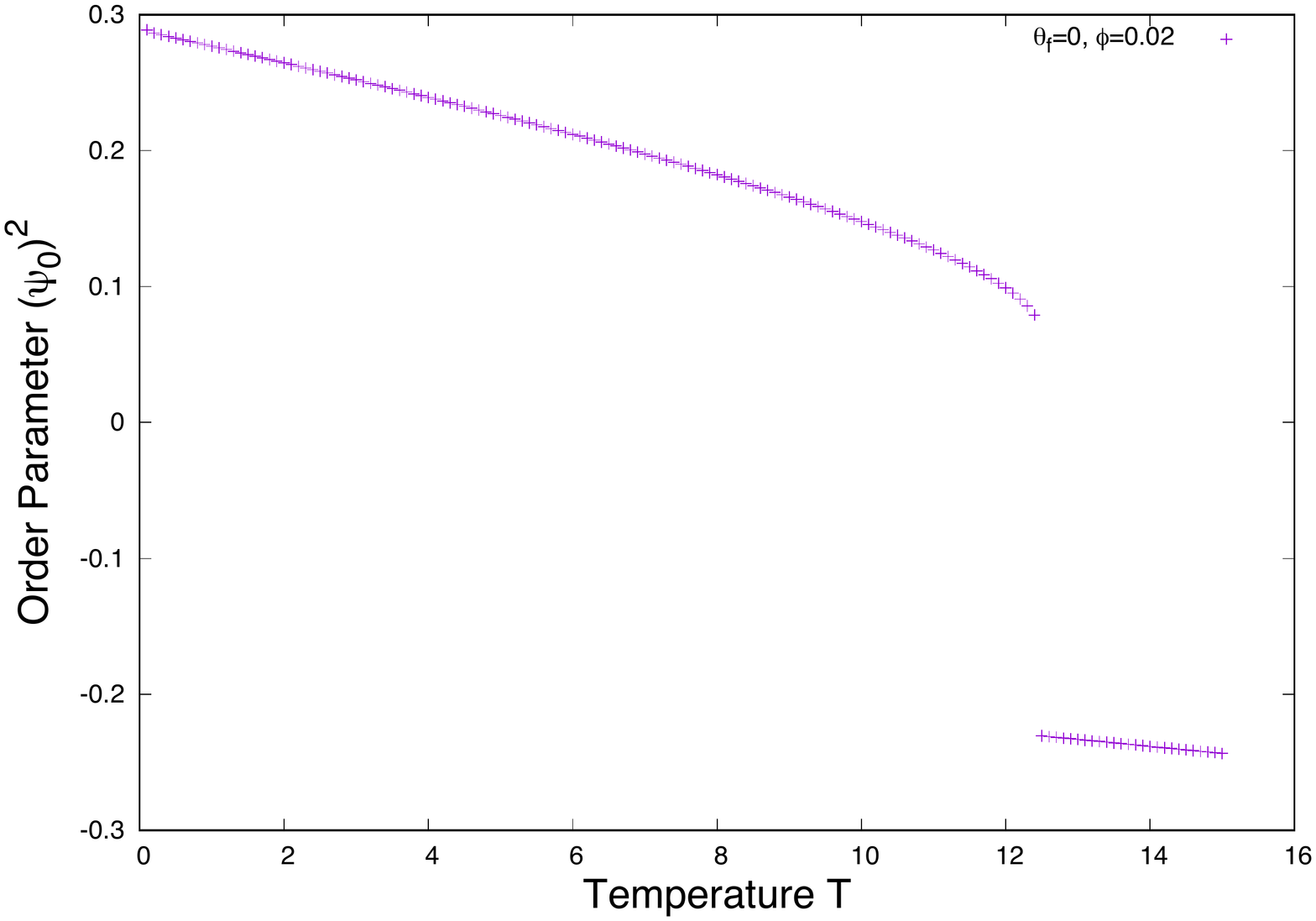}
     }
     \caption{$S$ and $\psi_0^2$ plotted against temperature $T$; show discontinuous jump for $\phi = 0.02$.}
     \label{fig3}
   \end{figure}

\begin{figure}[!ht]
     \subfloat[$S$ vs $T$ for $\theta_f=\pi/2$, $\phi=0.02$\label{subfig1:fig4}]{%
       \includegraphics[width=0.3\textwidth,angle=270]{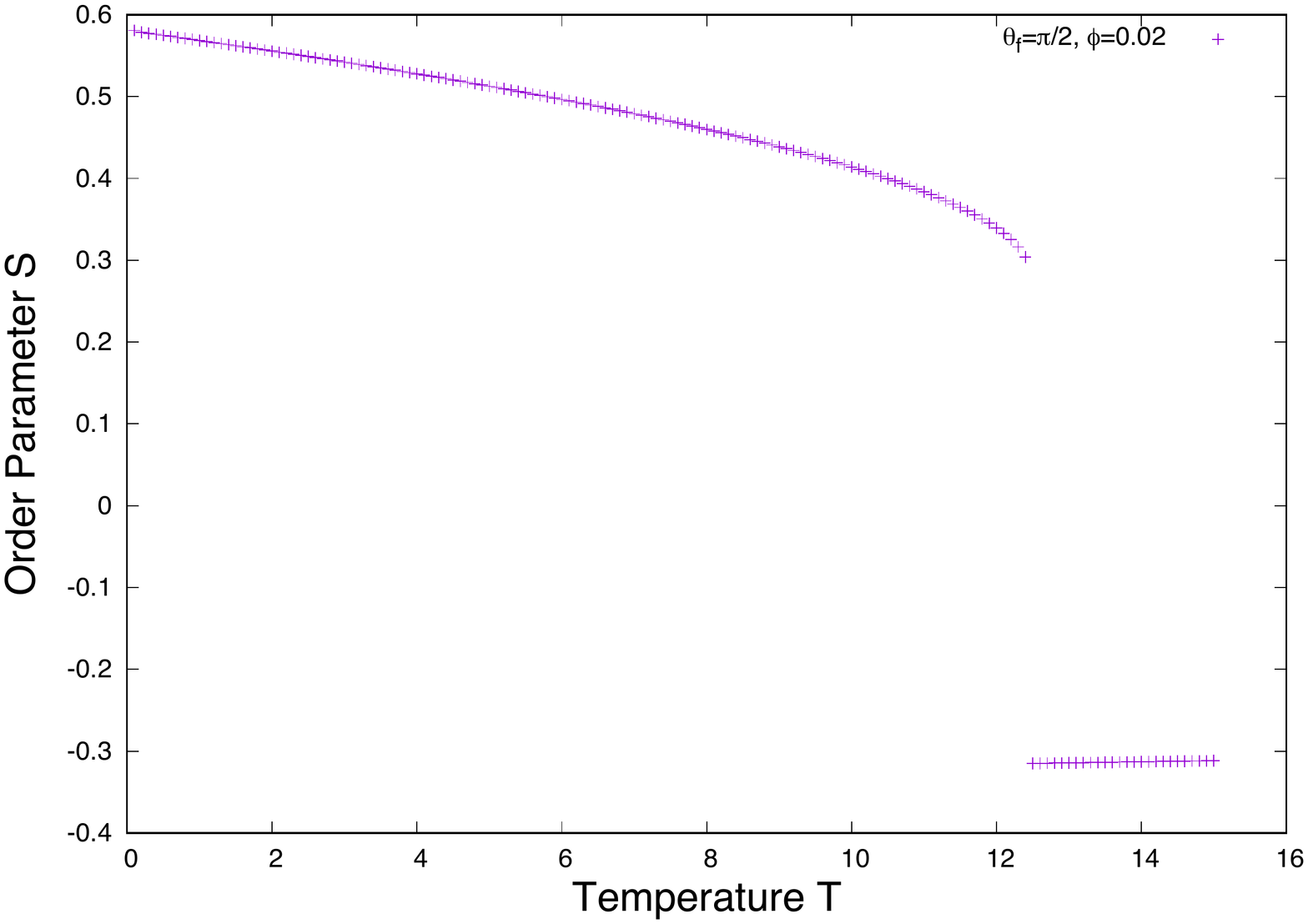}
     }
     \hfill
     \subfloat[$\psi_0^2$ vs $T$ for $\theta_f=\pi/2$, $\phi=0.02$\label{subfig2:fig4}]{%
       \includegraphics[width=0.3\textwidth,angle=270]{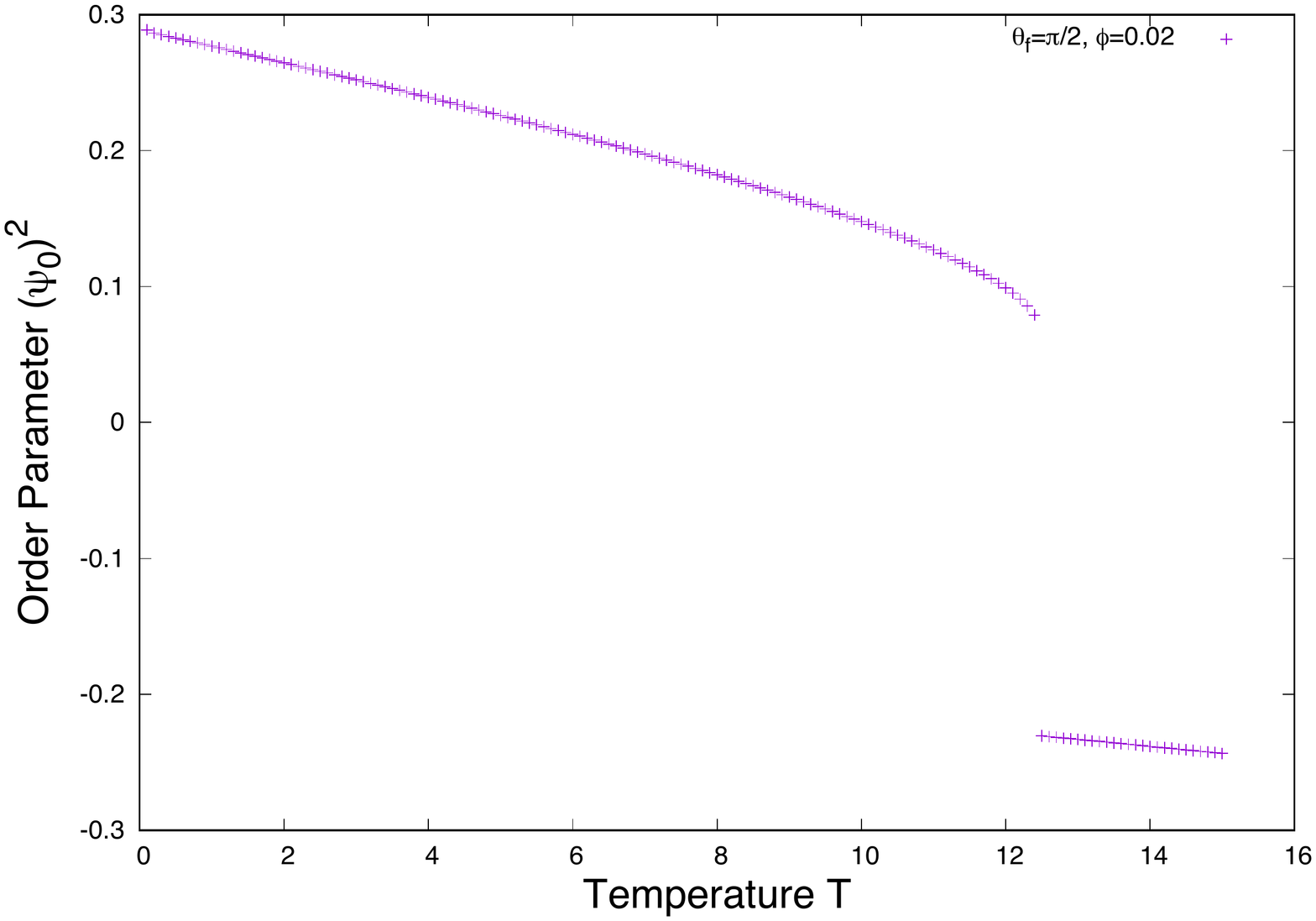}
     }
     \caption{$S$ and $\psi_0^2$ plotted against temperature $T$; show discontinuous jump for $\phi = 0.02$.}
     \label{fig4}
   \end{figure}

The temperature variations of the order parameters 
($S$ and $\psi_0$ ) for $\phi=0$ in  the 
FSmA phase are shown in Fig. \ref{fig1} and Fig. \ref{fig2}, respectively for $\theta_f=0,\pi/2$. 
Figs. \ref{fig3} and \ref{fig4} show the temperature variations of the same order parameters at the first order FN-FSmA transition point with the change of the concentration of FNP for $\phi \neq 0$ (= 0.02). Figs \ref{fig3} and \ref{fig4} show discontinuous phase transitions, as expected. 

For threshold temperatures $T_1=T_2=10$ (in non-dimensionalized units), we find continuous (second order) phase transition for the parameter values $\theta_f=0.0$ and $\phi=0.02$ (Fig. \ref{fig1}). However, changing the parameter values to $\theta_f=\pi/2$ and $\phi=0$ (Fig. \ref{fig2}) or $\phi=0.02$ (Fig. \ref{fig3}), we find discontinuous jumps, confirming our prediction of a first order phase transition.

The figures above have been drawn using the parameter values $a_0$= 0.012, $b=0.44$, $c=1.56$, 
$\alpha_0$=0.012, $p=0.1$, $q=2.0$, $\beta$ =2.07, 
$\gamma^{\prime}$= 2.98, $\eta^{\prime}$=-0.82, $\delta=1.01$ and $\omega=0.72$.  
The coefficients (parameters) entering
the Landau free energy are arbitrary and can at best be fitted to the experimental
data so as to give the observed physical behavior. Since enough experimental data
are not available in the literature, we cannot provide precise quantitative comparison of the values of the
coefficients used here against known experimental benchmark. Hence, we have restricted 
ourselves to generic parametric windows that portray the physically relevant behavior, focusing on the phase
transition aspect. Our choice of parameter values have been guided by the constraint of eliciting continuous
variations of $S$ and $\psi_0^2$ against $T$ for $\phi=0$ (second order)
while showing a discontinuous variation for $\phi \neq 0$ (first order). Generally, all we can say here is that the values of $\gamma^{\prime}, \omega>0$ while $\eta^{\prime}<0$.

The conditions for the first order FN-FSmA phase transition are given by
\begin{equation}
f_{FSmA}(S)=f_{FN}(S), f_{FSmA}^\prime (S)=0, f_{FSmA}^{\prime\prime}(S)\ge 0
\label{cond2}
\end{equation}
The conditions for the second order phase N-SmA transition read
\begin{equation}
\alpha^*-\delta^*S-\frac {\phi(1-\phi)\omega \eta^{\prime}}{q}S^2=0, f_{FN}^\prime (S)=0, f_{FSmA}^{\prime\prime}(S)\ge 0
\label{cond3}
\end{equation}
Again the conditions for phase equilibrium require that the chemical potentials
in the FN and FSmA phases are equivalent i.e $\mu_{FN}=\mu_{FSmA}$.

\begin{center}
\begin{figure}[ht]
\includegraphics[height=14.0cm,width=12.0cm,angle=270]{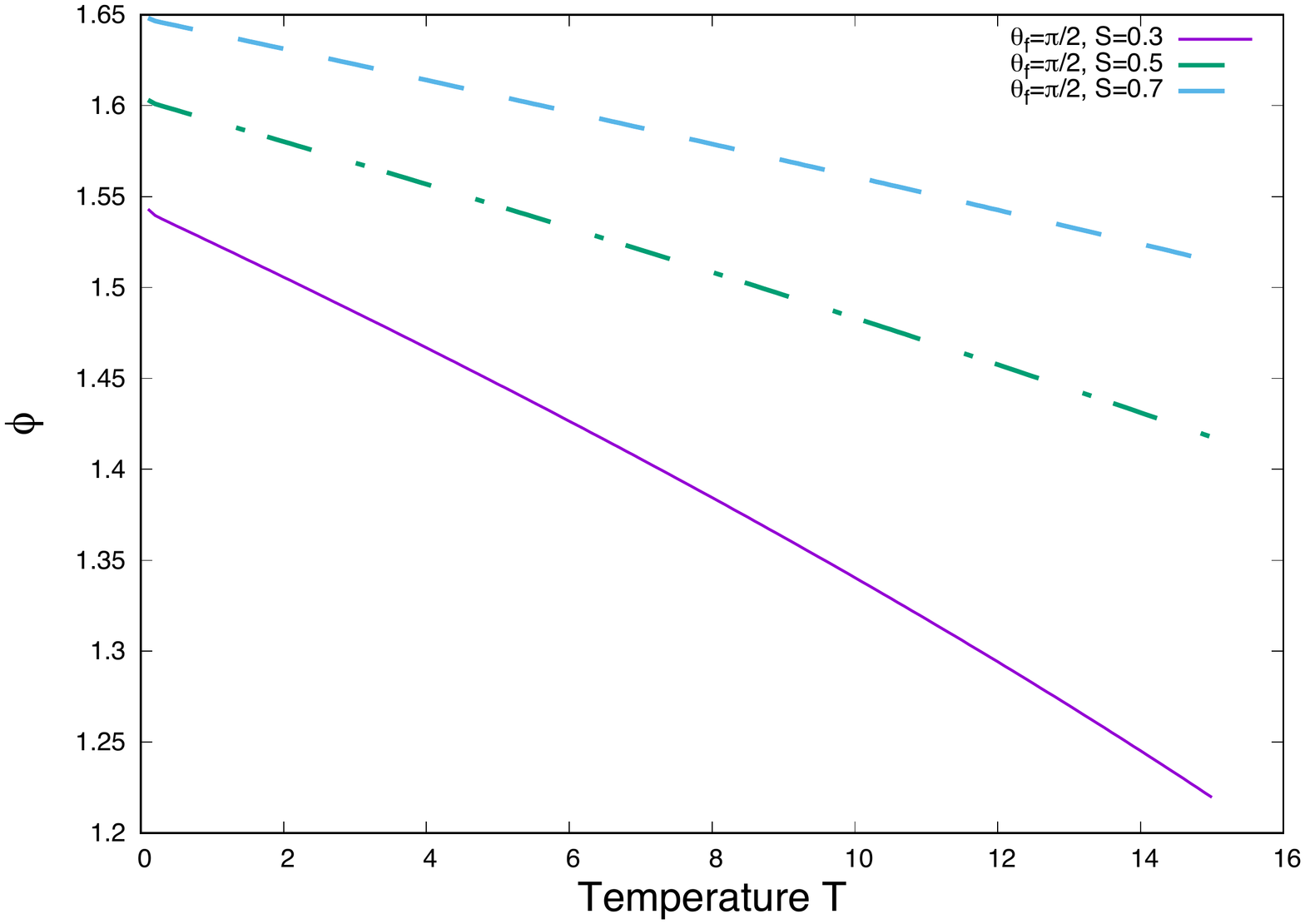}
\caption{$\phi$ plotted against temperature $T$ shows a continuous decay over a wide range of values of $S$ for $\theta_f=\pi/2$. The dot-dashed, dashed and solid lines respectively represent the cases for $S$ = 0.3, 0.5 and 0.7.
\label{fig5}}
\end{figure}
\end{center}

The $\phi$ versus temperature $T$ variation in the FSmA and FN phases, obtained from a self-consistent solution of Eq. (\ref{cond3}), is shown in Fig. \ref{fig5}. The results are shown over a range of values of $S$, including the phase transition value $S\sim 0.3$, as previously demonstrated in Figs. \ref{fig1}-\ref{fig4}. Apart from the generic decaying trend, the results remain largely unaffected by changes in $\theta_f$ values.

\section{Tricritical behavior of the FN-FSmA phase transition}

In this section we discuss the tricritical behavior of
the FN-FSmA phase transition under the influence of FNP.
Assume $S_0$ is the
order parameter of the FN phase at the FN-FSmA transition
point and $f_{FN}(S_0)$ is the corresponding free energy density at the
FN phase. Then the free energy density for a mixture of liquid crystal
and FNP near the FN-FSmA phase transition can
be written as
\begin{eqnarray}
f &=&(1-\phi)(f_{FN}(S_0)+\frac 12 u(S-S_0)^2
+\frac 12\alpha \psi_0^2 
+\frac 14\beta \psi _0^4-\frac 12\delta \psi_0^S)
+\phi(\frac 12p M^2
+\frac 14q M^4)\nonumber \\
&&-\phi(1-\phi)\left(\frac 14\gamma^{\prime} M^2S 
+\frac 12\eta^{\prime} M^2S^2
+\frac 12 \omega M^2\psi_0^2\right)
\label{free7}
\end{eqnarray}
where $F_{FN}(S_0)$ is the corresponding free energy density of the FN phase and
$u=1/\chi_1 $, $\chi_1 $ is the response function of the FN phase.

After eliminating the values of $S$ and $M$ from Eq. (\ref{free7}), we get the
free energy density as
\begin{equation}
F=F_{FN}^{*}(S_0)+\frac{1}{2}\alpha^{**}\psi_0^2+\frac{1}{4}\beta^{**}
\psi_0^4 
\label{free8}
\end{equation}
The renormalized coefficients are

$F_{FN}^{*}(S_0) =F_{FN}(S_0)
-\frac{p^{**2}}{4q^*}$,

$\alpha^{**}=\alpha-\delta S_0-\phi\frac {p^{**}\omega^{*}}{q^{*}}$,

$\beta^{**}=\beta-\frac {\delta^2}{2u}-\phi(1-\phi)\frac {\omega^{*2}}{q^*}$,

$p^{**}=p+\phi(1-\phi)\frac {\gamma^{\prime 2}}{4u}-(1-\phi)\gamma^{\prime}S_0-
(1-\phi)\eta^{\prime}S_0^2$,

$q^*=q-\phi(1-\phi)\frac {\gamma^{\prime 2}}{u}-\phi(1-\phi)\frac {2\eta^{\prime 2}}{u}S_0^2-\phi(1-\phi)\frac{\eta^{\prime}\gamma^{\prime}}{u}S_0$,

$\omega^*=\omega+\frac {\delta}{u}(\eta^{\prime}S_0+\gamma^{\prime})$.

It is clear from the renormalized coefficients that the parameters $\alpha^{**}$ and
$\beta^{**}$ change
with the change of concentration $\phi$ which indicates change of the order of the FN-FSmA phase
transition. For pure 8CB or low value of the concentration $\phi$ of FNP,
$\beta^{**}>0$, then a second order transition occurs.

Then renormalization of the second order FN-FSmA transition temperature 
can be written as
\begin{equation}
T_{NA}^{C}=T_{FN-FSmA}^*-\frac {\delta S_0}{\alpha_0^*}
-\frac {\omega^*}{q^*\alpha_0^*}\phi\left(\frac {\gamma^{\prime 2}}{4u}-
(1-\phi)\gamma^{\prime}S_0+(1-\phi)\eta^{\prime}S_0^2\right)
\label{stemp1}
\end{equation}
where 

where $T_{FN-FSmA}^*=\frac {\alpha_0T_2^*+\frac {p_0\omega^*T_f\phi}{q^*}}{\alpha_0^*}$,

$\alpha_0^*=\alpha_0+\frac {\phi\omega^*p_0}{q^*}$.

For the higher value of concentration $\phi$
of the FNP,
$\beta^{**}<0$, the FN-FSmA phase transition is a first order
transition.  In this case both the N and SmA phases can
coexist i.e. a two phase region appears. In this case sixth order term $\frac 
{e}{6}\phi_0^6$ should be added into the free energy density (\ref{free8}). 
Then for the first order FN-FSmA phase transition, the FN-FSmA transition temperature is
\begin{equation}
T_{FN-FSmA}=T_{FN-FSmA}^*+\frac {3\beta^{** 2}}{16e\alpha^*_0}-\frac {\delta S_0}{\alpha_0^*}
-\frac {\omega^*}{q^*\alpha_0^*}\phi\left(\frac {\gamma^{\prime 2}}{4u}-
(1-\phi)\gamma^{\prime}S_0+(1-\phi)\eta^{\prime}S_0^2\right)
\label{stemp2}
\end{equation}

Equations (\ref{stemp1}) and (\ref{stemp2}) show that the FN-FSmA transition
temperature decreases with increase of the concentration of FNP. This
prediction confirms the experimental results \cite{cordo3}.

For a tricritical value of the concentration $\phi_{tcp}$,  $\beta^{**}=0$, then a tricritical point is obtained. Hence
a TCP is achieved with the change of concentration of the FNP.

\section{Conclusions}

We have developed a phenomenological model combining with Flory-Huggins theory
to describe the effect of ferromagnetic nanoparticles on the I-N and N-SmA 
phase transitions. The I-FN transition is still a weakly first order 
transition even in the mixture of FNP. 
The FN-FSmA transition may be first order. In a binary mixture, the N-SmA
transition which is second
order only in one of the pure forms, becomes first order with the change of concentration of the FNP. This leads to a crossover from second to first order 
transition via TCP. Furthermore, both the I-FN and FN-FSmA transition 
temperatures decrease with the increase of the concentration of the FNP. 
We discuss our analysis by plotting various topology of the phase diagram 
under different conditions.
Our results are qualitative agreement with the experimental results.
Our results are expected to encourage further experiments on the impact of 
FNP on other 
liquid crystalline phase transitions to verify the validity of the present
theory.

In the present work we have discarded
spatial variations in the order parameter. The inclusion 
of these derivative terms will give additional
physics into these phase transitions.

\newpage
\noindent

\end{document}